# Scaffold Splits Overestimate Virtual Screening Performance

Qianrong Guo[a], Saiveth Hernandez-Hernandez[b], Pedro J. Ballester[a*1]

[a] Department of Bioengineering, Imperial College London, London SW7 2AZ, UK

[b] Centre de Recherche en Cancérologie de Marseille, Marseille 13009, France

**\* Corresponding author:** Pedro J. Ballester (p.ballester@imperial.ac.uk)

**Abstract.** Virtual Screening (VS) of vast compound libraries guided by Artificial Intelligence (AI) models is a highly productive approach to early drug discovery. Data splitting is crucial for better benchmarking of such AI models. Traditional random data splits produce similar molecules between training and test sets, conflicting with the reality of VS libraries which mostly contain structurally distinct compounds. Scaffold split, grouping molecules by shared core structure, is widely considered to reflect this real-world scenario. However, here we show that the scaffold split also overestimates VS performance. The reason is that molecules with different chemical scaffolds are often similar, which hence introduces unrealistically high similarities between training molecules and test molecules following a scaffold split. Our study examined three representative AI models on 60 NCI-60 datasets, each with approximately 30,000 to 50,000 molecules tested on a different cancer cell line. Each dataset was split with three methods: scaffold, Butina clustering and the more accurate Uniform Manifold Approximation and Projection (UMAP) clustering. Regardless of the model, model performance is much worse with UMAP splits from the results of the 2100 models trained and evaluated for each algorithm and split. These robust results demonstrate the need for more realistic data splits to tune, compare, and select models for VS. For the same reason, avoiding the scaffold split is also recommended for other molecular property prediction problems. The code to reproduce these results is available at https://github.com/ScaffoldSplitsOverestimateVS

## 1 Introduction

Virtual Screening (VS) has become a key drug discovery technique [28]. VS offers a fast and more cost-efficient method to discover novel molecules that have the potential to bind to a given target (structure-based VS) or have for instance growth inhibition activity on cancer cell lines (phenotype-based VS). Applying computational algorithms and models, VS can significantly reduce the time and financial burdens in the preclinical stages of drug discovery, where it accounts for over 43% of costs in the pharmaceutical

[1] Supported by CONAHCYT [S.H-H.] and the Wolfson Foundation and the Royal Society for a Royal Society Wolfson Fellowship [P.J.B.].

Q. Guo et al.industry [1]. Furthermore, VS permits screening gigascale chemical spaces [28] to enable novel drug discovery.

The application of Artificial Intelligence (AI) to improve VS and related problems has tremendous potential [2,25,9,10,16,23,7,40]. Many AI algorithms achieved great performance on popular benchmarks such as those in MoleculeNet [38]. In sharp contrast, AI models able to discover novel molecules with potent whole-cell activity are not being reported, with very few yet outstanding exceptions [22,37]. This paucity of successful prospective results is partly due to current benchmarks not capturing the challenges posed by real-world screening libraries. This gap highlights the need for improving benchmarks that more accurately represent the complexities of identifying potent drug leads.

Identifying the scaffold of a biologically active molecule is a common task in early drug discovery. These scaffolds form the foundation for drug development through chemical modification to enhance its medicinal properties [21]. Yet, the challenge lies in the subjective nature of determining what constitutes the core scaffold and vital side chains. This challenge emphasized the need for precise methodologies in scaffold identification to contribute to new drug development.

The chemical diversity challenge is one of the most important roadblocks to model generalization in drug discovery. Recent studies reported decreases in model prediction accuracy when applied to new drugs compared to new cell lines, illustrating the challenges in extrapolating across diverse chemical spaces [20,39,4,5,19].

To evaluate the model generalizability on dissimilar molecules, the scaffold split method has been proposed [38]. This method splits data instances (here molecules) into training, validation, and test sets based on the molecule scaffolds [3], offering a more challenging evaluation of model generalizability. For instance, Geometry-Enhanced molecular representation learning Method (GEM) obtained top performance on MoleculeNet against other methods using the scaffold split [6].

While the scaffold split is often regarded as realistic (e.g. [36]), we argue that this approach does not accurately reflect the real chemical diversity in the compound library to the models. In the vastly diverse libraries such as ZINC [14] and limited scaffolds represented in the training sets, scaffold split will be less representative due to the scaffolds in the training set less (even by orders of magnitude) than the scaffolds in ZINC20. While scaffold splits aim to enhance the model predictivity for dissimilar chemicals, their ability to simulate real-world situations remains debatable.

Our analysis demonstrates that the commonly used scaffold split is not realistic in that molecules with different chemical scaffolds are often similar. Surprisingly, this key observation has been overlooked since this split was introduced [38]. Here we use splits based on Uniform Manifold Approximation and Projection (UMAP) [11], which introduces more realistic distribution shifts than scaffold splits. Furthermore, ROC AUC used as the primary metric in MoleculeNet is a suboptimal metric for VS, as it does not focus on the early-recognition performance. The primary metric in VS should instead be hit rate or similar early-recognition metrics, which helps to avoid mistaking inactive molecules as inactives. To investigate these issues, we employ a benchmark with extensive labeled data across 60 cell lines. Thereby reducing the dataset size as a limiting factor for model performance evaluation in testing our hypothesis against the backdrop of chemical diversity and model generalizability.

## 2 Experimental design

### 2.1 Datasets

Our study utilized data from the US National Cancer Institute's NCI-60 cell line screening program [30], released in Oct 2020[2]. The datasets comprise data of 60 cell lines from nine tumor types: renal cancer, prostate cancer, ovarian cancer, brain cancer, colon cancer, non-small-cell lung cancer, melanoma, breast cancer, and leukemia, to help identify novel anticancer drugs. The growth inhibition conducted by a molecule on a given cell line is summarized by the $GI_{50}$ data for all 60 cell lines. $GI_{50}$ represents the concentration required to inhibit cell growth by 50%, serving as a key indicator of a compound's potential anticancer activity. The optimal clustering of these molecules [11] revealed seven clusters comprising 33, 118 unique molecules across all 60 cell line datasets with 1, 764, 938 $GI_{50}$ determinations (88.8% completeness). We operate with $pGI_{50}$ (the negative base 10 logarithm of $GI_{50}$ in molar units). Thus, a higher $pGI_{50}$ value corresponds to higher inhibitory potency.

### 2.2 Scaffold split for each of the 60 datasets

The widely used scaffold split [38] is implemented in RDKit (https://www.rdkit.org/) using the Bemis-Murcko method for scaffold identification. Compared to random splits, scaffold splits are pointed out to be more challenging for models evaluated in chemically diverse datasets [35]. To match the number of clusters used across different splitting methods, we divide the data into seven groups using the scaffold split such that molecules sharing the same scaffold are grouped together. This partition was performed by evenly distributing a total of 33,118 unique molecules into seven groups, resulting in an approximate group size of 4,731.

### 2.3 UMAP and Butina clustering splits for each of the 60 datasets

We employed the UMAP clustering of NCI-60 molecules found optimal in a previous study [11]. In a nutshell, UMAP was employed to reduce the dimensionality of the molecular Morgan fingerprints and clustering was performed on the molecules based on their reduced dimensions. We found that 7 UMAP clusters of these molecules were of higher quality than those provided by Butina clustering and Hierarchical clustering. Here we define the UMAP split as taking one of these seven clusters as a test set and the rest merged for the training set. Since the number of molecules differs across clusters in the UMAP split, we chose the cluster of 4,396 molecules as the test set for the UMAP split for single cell line comparison, which has the closest number of molecules as the test set size in the scaffold split. Lastly, for a given cell line, only the molecules with an annotated $pGI_{50}$ for that cell line can and are retained in the corresponding data split. For each cell line, this procedure ensures that the training set and test set do not only contain different molecules, but also that these have dissimilar chemical structures. The Butina split of dataset is made from the Butina clustering[11] in a completely analogous manner than it was just explained for UMAP clustering.

---

[2] The URL for downloading this version of NCI-60 data is:
https://wiki.nci.nih.gov/display/NCIDTPdata/NCI-60+Data+Download+-+Previous+Releases

## 2.4 Comparative evaluation across the 60 datasets

To evaluate the model efficacy, we performed a seven-fold cross-validation, with 5 different seeds for each of the 60 datasets. This cross-validation employed the divided seven scaffold groups or seven molecule clusters, where each cross-validation fold used one group or cluster as a test set and the others for training.

## 2.5 Learning algorithms

We employed three models for comparison: the linear model Linear Regression (LR), the tree-based model Random Forest (RF), and the pre-trained graph-based deep neural network model (GEM).

LR [8] linearly models the relationships between the pre-calculated features and $pGI_{50}$ as the label to predict. The LR model was built using the Python package scikit-learn [27].

RF is a supervised learning algorithm building an ensemble of simple decision trees [12]. When building one decision tree for the RF algorithm, we first pick a subset of the same number of features as the original dataset, with the only difference being that it may include repeated data from the original dataset (bootstrapping). Then, at each training step, the random subset is used as the training set. By repeating the above-described procedure, we will have a variety of decision trees forming an RF model. The RF model makes the predictions by averaging the predicted $pGI_{50}$. This approach that applies bootstrapping and aggregation for predicting is called "bagging". In this study, we built the RF model for the regression task using the Python package scikit-learn [27].

In addition to these two well-established algorithms, we also used the pre-trained regression model GEM [6], a novel approach that utilizes both molecular topology and geometry information for molecular representation learning. GEM adapts a geometry-based Graph Neural Network (GeoGNN) model pre-trained on 20 million unlabeled molecules from ZINC15 [32], which considers atoms, bonds, and bond angles in a molecule simultaneously. In our study, we loaded the pre-trained GeoGNN parameters[3] and then fine-tuned the model for our downstream $pGI_{50}$ prediction task.

The pre-trained GeoGNN model is a graph neural network model used in GEM. It takes in an atom bond graph and a bond angle graph as inputs and outputs the node representation, edge representation, and graph representation. The GeoGNN model consists of several layers of blocks, each of which contains a graph isomorphism network layer, a layer normalization layer, a graph normalization layer, a ReLU activation layer (if specified), and a dropout layer. In each of the blocks, the graph features will be updated through the aggregation process. If *u* and *v* are atoms and *(u, v)* the covalent bonds that connects them, during the model aggregation in the graph, at the *k-th* iteration, the vectors representing the node and edge are $h_u$ and $h_{uv}$. $AGGREGATE^{(k)}$ is the aggregation function in the graph neural networks, which can operate each node neighborhood before using the $COMBINE^{(k)}$ to combine all the updated node representations. The neighborhood atoms of the atom *u* are represented as *N(u)*. The GeoGNN block can be formulated as (adapted from [6]):

---

[3] The pre-trained parameters can be found at
https://baidunlp.bj.bcebos.com/PaddleHelix/pretrained_models/compound/pretrain_models-chemrl_gem.tgz

$$\alpha_{uv}^{(k)} = AGGREGATE_{bond-angle}^{(k)} \left(\left\{\left(h_{uv}^{(k-1)}, h_{uw}^{(k-1)}, x_{wuv}\right), w \in N(u)\right\}\right)$$

$$\cup \left\{\left(h_{uv}^{(k-1)}, h_{uw}^{(k-1)}, x_{wuv}\right), w \in N(u)\right\} \quad (1)$$

$$h_{uv}^{(k)} = COMBINE_{bond-angle}^{(k)}\left(h_{uv}^{(k-1)}, \alpha_{uv}^{(k)}\right) \quad (2)$$

The GeoGNN model uses mean pooling as the readout layer to obtain the graph representation. The graph pooling layer takes in the atom bond graph, the node representation, and the edge representation as inputs and outputs the final graph representation. After the GeoGNN, it is followed by a Multilayer Perceptron (MLP), which takes the graph representation as input and predicted pGI$_{50}$.

To predict these pGI$_{50}$ values, two types of input features were employed:
1. **Pre-calculated Features.** A total of 263 features comprising a Morgan fingerprint (256 bits, radius 2) plus 7 molecular physicochemical descriptors calculated using RDKit [17,18]. These features were used as input when applying the LR and RF algorithms.
2. **Molecular Graphs.** Molecular graphs that were constructed from the SMILES strings of the compounds were the input data for GEM [6].

## 2.6 Test set performance metrics

We assessed model performance using a dual regression-classification approach. The pGI$_{50}$ values larger than 6 (GI$_{50}$ = 1 μM) were considered as positive and below or equal 6 as negative. This approach allowed us to categorize each prediction as either True Positive (TP), True Negative (TN), False Positive (FP), or False Negative (FN). Evaluation metrics such as False Positive Rate (FPR) and True Positive Rate (TPR or Sensitivity) were calculated from them. Metrics for VS have been explained elsewhere [33], but for convenience are summarized below.

Hit rate, our primary metric, measures the proportion of TP in compounds identified as positive, as in Equation 3, which perfectly aligns with the objective of VS: screen vast compound libraries to identify potential leads. With a high hit rate, we are more confident in having more efficient results to find promising leads from a pool of candidates.

$$hit\ rate\ (\%) = \frac{TP}{TP + FP} \times 100\%$$

Matthews Correlation Coefficient (MCC) is a metric summarizing both types of errors:

$$MCC = \frac{TP \times TN - FP \times FN}{\sqrt{(TP + FP)(TP + FN)(TN + FP)(TN + FN)}}$$

The Receiver Operating Characteristic (ROC) curve illustrates the performance of a binary classifier as its discrimination threshold varies. It is the plot of TPR against FPR. TPR indicates the proportion of TPs being correctly identified as such, while FPR measures the proportion of negatives being incorrectly classified as positives. Thus, the Area Under the ROC Curve (AUC) evaluates the model's ability to distinguish between the classes, which in this case are active and inactive compounds. A higher ROC AUC value indicates better model performance, with a value of 1.0 representing a perfect classifier, and a value of 0.5 indicates no discriminative power (equivalent to random guessing).

Root Mean Square Error (RMSE) is the square root of the mean of the squared differences between the actual pGI50 $(y_i)$ and the model predicted pGI50 $(\hat{y}_i)$:

$$RMSE = \sqrt{\frac{\sum_{i=1}^{n}(y_i - \hat{y}_i)^2}{n}}$$

## 2.7 Model hyperparameters

We used the RF implementation provided by the scikit-learn Python package [27]. The model was trained using the default settings, with the hyperparameters provided in Table 1. As for GEM, we used the pre-trained GeoGNN model parameters provided by its authors. We fine-tuned the GEM for 20 epochs on our dataset, using the default settings, including a batch size of 32, a dropout rate of 0.1, and a learning rate of 0.001. The hyperparameters used are shown in Table 1. We chose the default hyperparameters to ensure that the model could be easily compared to other models.

**Table 1.** Hyperparameters for RF and GEM. The head learning rate is the learning rate for the MLP after the pre-trained GeoGNN, and the encoder learning rate is the learning rate for the pre-trained GeoGNN.

| Algorithm | Hyperparameter | Default Value |
|---|---|---|
| RF | n_estimator | 100 |
| | criterion | squared error |
| | max_depth | None |
| | min samples split | 2 |
| | min samples leaf | 1 |
| | min weight fraction leaf | 0.0 |
| | max features | 1.0 |
| | max leaf nodes | None |
| | min impurity decrease | 0.0 |
| | bootstrap | True |
| | max samples | None |
| GEM | Batch Size | 32 |
| | Dropout Rate | 0.1 |
| | Encoder Learning Rate | 0.001 |
| | Head Learning Rate | 0.001 |
| | Epoch | 100 |

# 3 Results

## 3.1 Limitations of scaffold split to generate realistic VS benchmarks

The scaffold split ensures that the test set only contains molecules with unseen scaffolds. That is, there are no training molecules with any of the scaffolds in the test molecules. What has not been noted yet is that scaffolds are often similar and can be almost identical. Take, for instance, the 48, 416 molecules tested on the IGROV1 cell line by the NCI-60. The two most frequent scaffolds in these molecules are benzene and pyridine (Fig. 1A), which are almost identical. The scaffold split can place all benzene-containing molecules in one set and all pyridine-containing molecules in the other set. Thus, high training-test similarities are introduced by the scaffold split (Fig. 1B).

The resulting benchmarks will hence not be realistic, as the real test sets, compound libraries for prospective VS, contain not only new, but novel scaffolds [14]. Evaluating models using the scaffold split overestimates their prospective performance, as those real test sets are more complex and challenging. Thus, current benchmarks can miss the most promising models for prospective applications. There is hence an urgent need to investigate the impact of this shortcoming and provide data splits that are closer to the real level of difficulty posed by prospective VS.

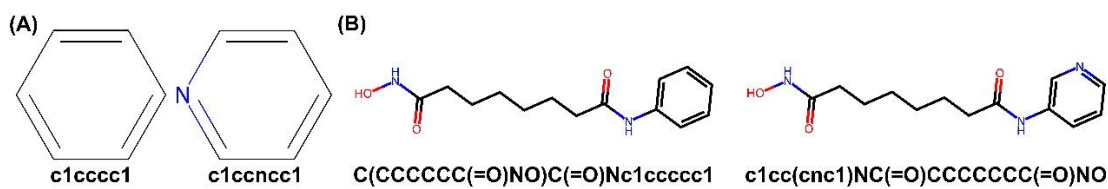

**Fig. 1. Bemis-Murcko scaffolds can be almost identical, and thus the scaffold split often introduces strong training-test similarities.** (A) Benzene and pyridine are the two most frequent scaffolds in the molecules tested on the NCI-60 IGROV1 cell line (each scaffold has its own SMILES string underneath). The scaffold split ensures that the test set only contains molecules with unseen scaffolds, i.e., no molecules in the training set have the same scaffold in the test set. (B) However, there will be molecules with similar scaffolds, which would not all be placed in the same set despite being globally similar, e.g., benzene-containing molecules could be assigned to the test set when pyridine-containing molecules are in the training set. The most similar pairs of molecules with those different scaffolds are shown. As a result, the performance of a model trained and evaluated on the scaffold split will be overestimated.

## 3.2 GEM and RF perform similarly with scaffold split

We first started by comparing GEM to a nonlinear baseline (RF) and a linear baseline (LR). We adopted a dual regression-classification evaluation approach, wherein molecules were classified as positive and negative based on their activity cutoff at $pGI_{50} = 6$ (1 μM $GI_{50}$). In Fig. 2, we explored the relationship between predicted and measured $pGI_{50}$ values for molecules using scaffold split. We selected the IGROV1 cell line for this initial analysis as its results are representative of those from the other cell lines. The scaffold split was employed to train and test LR, RF, and GEM.

Q. Guo et al.

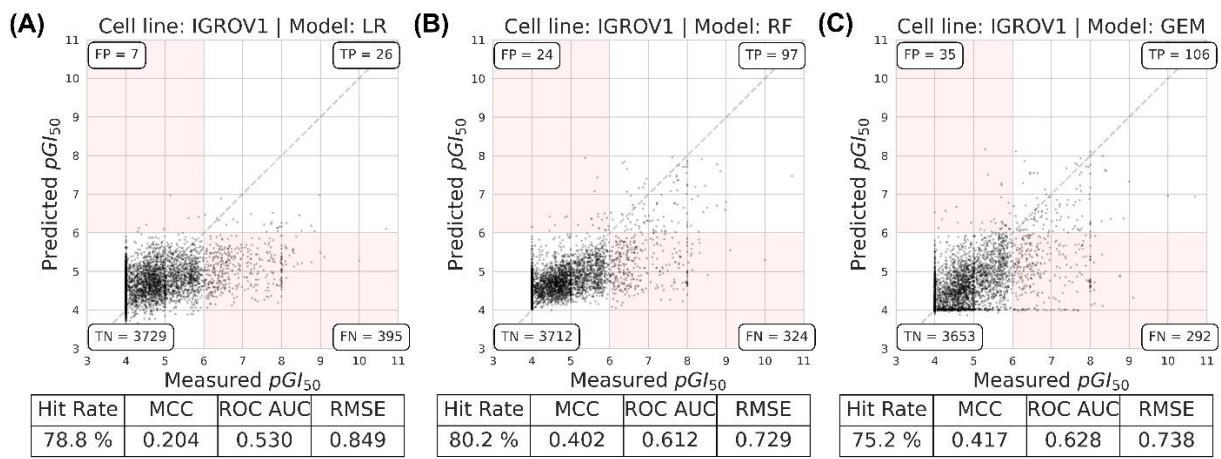

**Fig. 2. Predicted pGI$_{50}$ vs Measured pGI$_{50}$ of molecules on the IGROV1 cell line using the scaffold split with different regression algorithms.** (A) LR. (B) RF. (C) GEM. These results are from the first run (seed 1 of 5) and first partition (fold 1 of 7) of the molecules with their pGI$_{50}$s measured on the IGROV1 cell line. The test set contains 4,086 of these molecules, with the rest in the IGROV1 dataset used to train the three models (they the share the same training set). Note that, while constructing molecule graphs, 476 molecules from the training set and 71 molecules from the test set failed to generate the graph and were removed from the dataset for GEM.

Based on the evaluation results, RF would be the preferred model for the highest hit rate among these three models for prospective VS against the IGROV1 cell line. Even though GEM is the more complex method with the best MCC and ROC AUC scores, GEM obtained the lowest hit rate, which is the most informative metric for VS. Additionally, as GEM obtains a ROC AUC of 0.628 on IGROV1 (Fig. 2), this benchmark is much more demanding than the three using scaffold splits in MoleculeNet [38], where GEM obtains ROC AUCs of 0.806 (HIV), 0.856 (BACE), and 0.724 (BBBP) [6].

**3.3 GEM outperforms RF when using the more realistic UMAP clustering split**

Fig. 3 (bottom row) shows the results with the UMAP clustering split, which are much worse than with the overestimated scaffold split. LR and RF both exhibit a massive drop in performance metrics, from a hit rate of about 80% to now 0%. While GEM also performed much worse on this split, it is the only model to achieve a non-zero hit rate (11.9%) on this run using the IGROV1 dataset. Fig. 3 (top row) shows the results with the Butina clustering split. The results show that the Butina split results in a substantially easier benchmark than the UMAP split, consistent with the quality of its clustering being also lower on this dataset [11].

Compared to the scaffold split on the same dataset (Fig. 2), both clustering splits lead to a more demanding benchmark for this problem. These results highlight the importance of selecting the right computational tool for VS. The generalizability and predictability of GEM in identifying active compounds in a challenging and realistic dataset suggest that it is better able to retrieve inhibitors with novel chemical structures for the IGROV1 cell line in prospective settings.

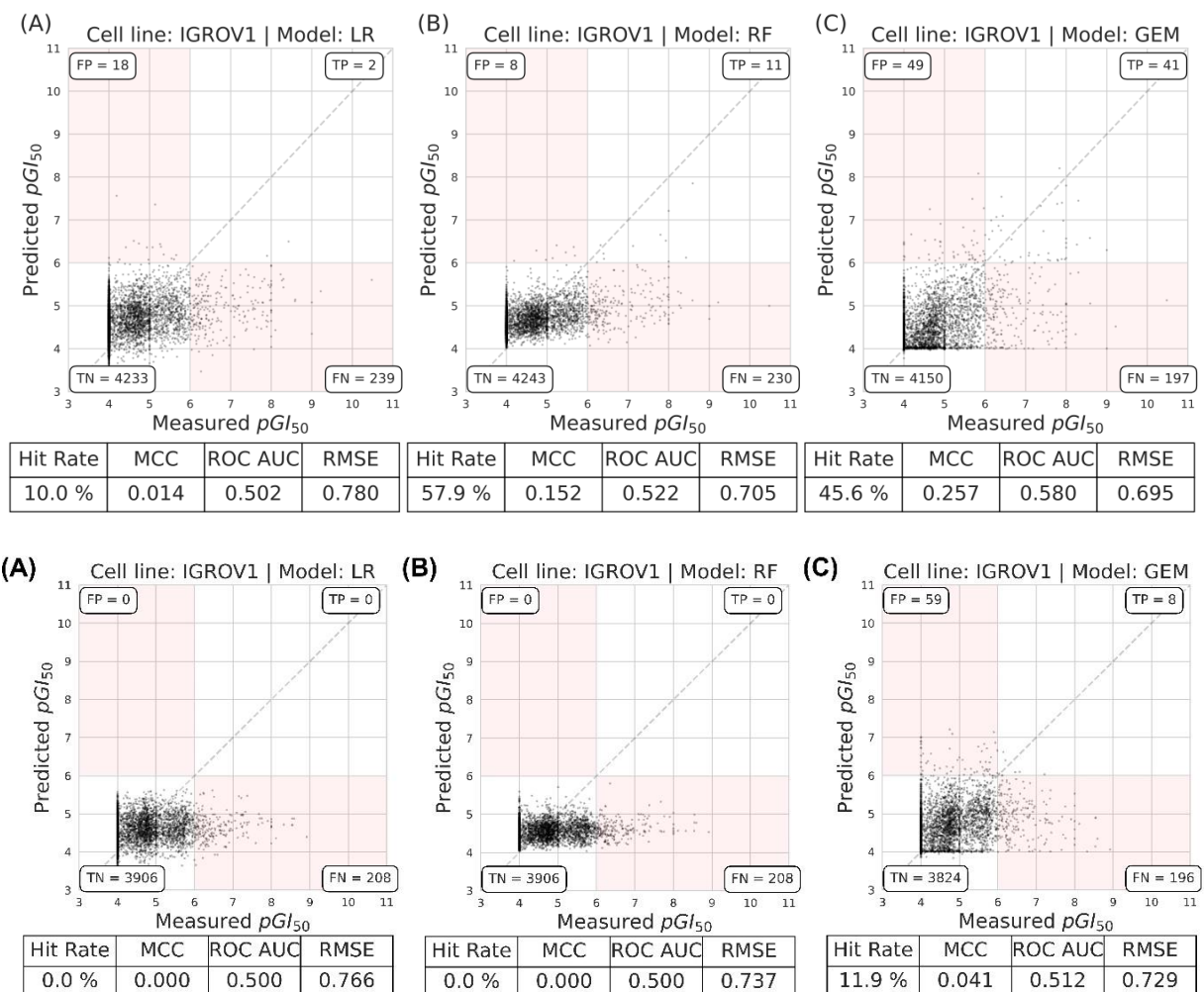

**Fig. 3. Predicted pGI$_{50}$ vs measured pGI$_{50}$ of molecules on the IGROV1 cell line using clustering splits with three different regression algorithms.** Butina split in the top row: (A) LR. (B) RF. (C) GEM. UMAP split in the bottom row: (A) LR. (B) RF. (C) GEM. Here we use the same dataset used for Fig. 2, simply split differently into training set and test set.

### 3.4 These results are robust across the 60 datasets

Similar trends to those observed with the IGROV1 cell line were observed in the evaluation across all the 60 datasets as illustrated in Fig. 4. In particular, with the scaffold split, RF is preferable as it has a higher median hit rate than the other two models (this happens even with the Butina split). Despite GEM performing better in terms of the median ROC AUC score, the hit rate is deemed a much more relevant metric due to intuitively evaluating for false positives (these have the highest priority in practical VS applications). Conversely, the UMAP split assessment, which is more realistic than those based on the scaffold split or even the Butina split, revealed that GEM strongly outperformed RF in both median hit rate and median MCC. This indicates that GEM will have increased efficacy in more realistic VS scenarios.

Note that both splits, scaffold and Butina, mislead model selection. The results with either of these splits would have led to the incorrect conclusion that RF was the optimal model based on the hit rate performance. However, UMAP split results show that the GEM model is a more suitable choice. These results corroborate

Q. Guo et al.

our hypothesis that scaffold split overestimate model performance and is misleading in selecting models for prospective use. This has also been the case for the Butina split, although to a lesser extent.

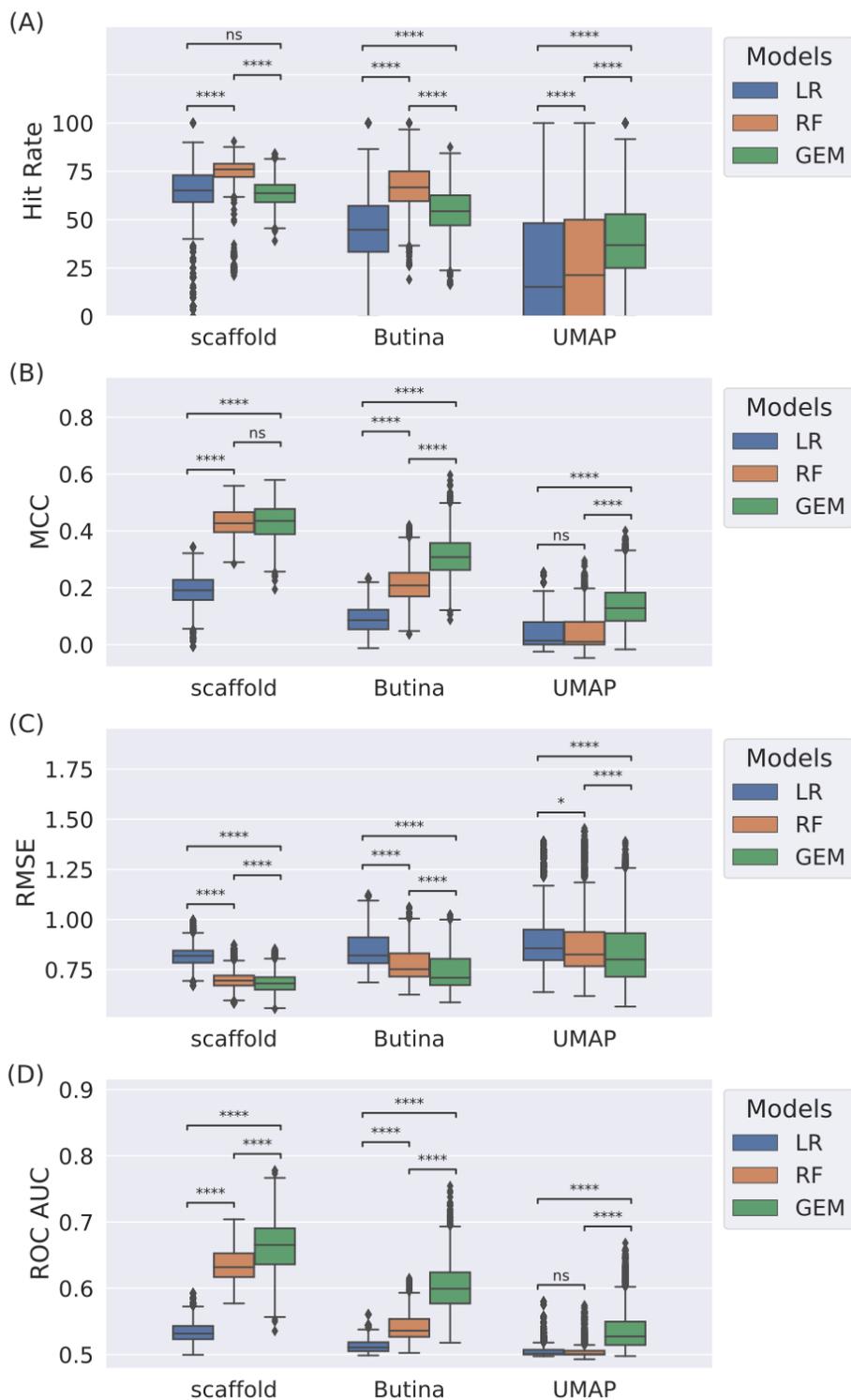

**Fig. 4. Cross-validation results for the three regression algorithms across the 7 left-out folds, 60 datasets and 5 seeds with each splitting method.** Out-of-sample performance was evaluated with four metrics: (A) Hit rate. (B) MCC. (C) RMSE. (D) ROC AUC. P-value (p) annotation legend: ns ($5.00 \times 10^{-2} < p \leq 1.00 \times 10^{0}$), * ($1.00 \times 10^{-2} < p \leq 5.00 \times 10^{-2}$), ** ($1.00 \times 10^{-3} < p \leq 1.00 \times 10^{-2}$), *** ($1.00 \times 10^{-4} < p \leq 1.00 \times 10^{-3}$), **** ($p \leq 1.00 \times 10^{-4}$).

It is worth noting how solid and reliable these model performance results are. Differences are highly significant for the discussed cases (pairs of boxplots), with strong effect size differences too. Each of the boxplots in Fig. 4 are formed by 5 x 7 x 60 = 2100 evaluations of the considered metric (one evaluation per trained model). That is, out-of-sample predictions for the molecules in the 7 left-out folds with each of the 60 datasets, repeated 5 times, each time initialized with a different random seed.

## 4 Discussion and Conclusion

The common misconception of scaffold split being able to mimic performance in real-world scenarios has been explained and demonstrated. The most ambitious objective of VS is not merely to identify active molecules with unseen scaffolds, but to discover those with dissimilar scaffolds. Identifying such a genuinely novel scaffolds is important for discovering new therapeutic pathways and overcoming drug resistance. The scaffolds in one group are strictly not the same as those in the other groups, but the molecules that contain them can exhibit extremely high similarities. In real-world VS, the molecules, including their scaffolds, can be very dissimilar within the chemical space. Thus, the UMAP clustering split method simulates better this real-world situation. By effectively clustering molecules based on their chemical similarity, we ensure that dissimilar molecules are assigned to different sets to evaluate the model's generalizability, which is more aligned with the goal of patentable drugs with novel scaffolds.

We prioritized hit rate as the primary metric for VS, aiming to reduce false positive errors in the early drug discovery stage, which are critical for identifying potential drug leads. Furthermore, we consider MCC as the secondary metric. Unlike hit rate that only captures FP errors, MCC measures both FN and FP errors. Hit rate and MCC are calculated with a cutoff, which in this study is $pGI_{50} > 6$ as positive (hence $pGI_{50} \leq 6$ as negative instance), while the most reported performances in machine learning literature are ROC AUC for classification and RMSE for regression tasks. However, these metrics are not as suitable for VS as the hit rate and MCC, as VS is a strongly class-imbalanced early recognition problem [34].

When using the scaffold split, the GEM's median ROC AUC score across the 60 problem instances outperforms that of RF. In contrast, when using early recognition metrics, specifically hit rate, the scaffold-split median performance of GEM is slightly worse than RF. When moving on to the UMAP clustering split, GEM's median performances in terms of ROC AUC and hit rate surpass those of RF. However, note that: a) GEM performance was much more modest, and it declined with scaffold splitting, suggesting an overestimation, although significantly better than that of RF or LR. b) The performance of RF is indistinguishable from that of LR, with both median performances being at a random level. These findings not only show the limitations of LR and RF models for prospective VS against cancer cell lines, but also highlight the capabilities of deep learning models to generalize to dissimilar molecules in this problem.

In conclusion, all the algorithms were not performing as well as they were expected. The application of the UMAP split, however, showed the true efficacy of the model in real-life scenarios, showing the reasons behind the limited success of AI in the drug discovery field and its ability to identify potential drugs.

Furthermore, situating our work within the broader field, specifically for the NCI-60 dataset, reveals a gap in research focused on unseen molecule prediction. Not many studies focus on unseen molecule prediction as our study but focus on unseen cell line prediction [4]. In our study, we evaluated the model on unseen molecules, which is fundamentally different from those studies and not directly comparable to studies

Q. Guo et al.across different cell lines [15]. Moreover, there are a few studies in the same domain predicting drug responses using NCI-60 data, these studies primarily relied on random splits [31,24] or at most scaffold splits [26] or did not research further to make predictions [29]. Our new approach, employing the UMAP clustering split permits evaluating the generalizability problem in drug discovery not just how the model performs across the biological context (i.e., on different cell lines), but also how the machine learning model can make robust and reliable drug response predictions of novel drug-like chemicals.

Our study has revealed the issue of the commonly used scaffold split overestimating model performance. Following proof-of-concept work showing that graph neural networks are particularly suited for VS against cancer cell lines [41]. We have demonstrated here the important limitations of employing the scaffold split on VS problems. As scaffold-split data introduces strong training-test similarities regardless of the label to predict, we also expect this split to overestimate model performance in molecular property prediction problems other than VS. This issue will be particularly acute with ultra-large libraries, which can potentially be larger than $10^{20}$ make-on-demand molecules [13]. Note that labeled molecules mostly come from in-stock libraries, the standard way to acquire them until recently [13], prior to *in vitro* experiments determining the label of interest (whole-cell activity in this study). Unsurprisingly, over 97% of the Bemis-Murcko scaffolds in make-on-demand libraries were already unavailable from in-stock libraries four years ago [14]. With fast growing chemical diversity, it is urgent to stop the misleading practice of using the scaffold split to evaluate molecular property prediction models.